# Extended elastic model for flow in metallic glasses

J. Q. Wang, W. H. Wang*, and H. Y. Bai*

*Institute of Physics, Chinese Academy of Sciences, Beijing 100190, P. R. China*

ABSTRACT

We report that both shear and bulk moduli, not only shear modulus, are critical parameters involved in both homogeneous and inhomogeneous flows in metallic glass. The flow activation energy ($\Delta F$) of various glasses when scaled with average molar volume $V_m$, which is defined as flow activation energy density $\rho_E = \Delta F/V_m$, can be expressed as: $\rho_E = \frac{10G+K}{11}$ . The extended elastic model is suggestive for understanding the glass transition and deformation in metallic glasses.

PACS：61.43.Fs, 81.05.Kf, 64.70.Pf

*To whom correspondence should be addressed. Email: whw@aphy.iphy.ac.cn and hybai@aphy.iphy.ac.cn

The mysterious glass transition phenomenon, which connects the liquid and glassy state, has wide applications in daily life, industry, and organism preservation [1-4]. In the past decades, intensive efforts have been made to understand the glass transition [1, 5-9]. To understand the flow in supercooled liquid and glass, many models have been proposed. The well-known models are the free volume model, the Adam-Gibbs entropy model, the mode-coupling theory and elastic models [1-6]. A successful model of viscous liquids and glasses must explain why the activation energy has such strong temperature dependence and can correlate the activation energy to simple and readily measurable parameter. Among these models, the elastic models link the glass transition and elastic moduli of the glasses [1]. All the elastic models link the activation energy to the readily measurable instantaneous shear modulus $G$. In metallic glasses, the glass transition temperature ($T_g$) indeed shows clear correlation with the elastic moduli such as Young's modulus $E$ and $G$ [10-18].

In this letter, based on the scaling laws between $T_g$ and elastic moduli in metallic glasses, we demonstrate that the $V_m$ scaled flow activation energy ($\Delta F$), that is, flow activation energy density, $\rho_E = \Delta F/V_m$, is determined by both $G$ and $K$ in a way of $\rho_E = (10G+K)/11$. The physical origin of the extended elastic model is discussed.

The temperature dependence of the viscosity of liquids approaching glass transition is [1]: $\eta = \eta_0 \exp\left[\frac{\Delta F(T)}{k_B T}\right]$, where, $\eta_0$ is a constant, $k_B$ is the Plank constant. At $T_g$, the viscosity of various liquids get to $\eta(T_g)=10^{13}$ poise [3] for metallic glasses. Then, the $\Delta F/T_g$ should be the same for metallic glasses. A variety of elastic models have been proposed [1], which assume that $\Delta F$ is proportional to the elastic modulus [1], e.g. $\Delta F \propto G$. Figure 1(a) shows experimental data of $E$ and $G$ versus $T_g$ for 46 different kinds of bulk metallic glasses (BMGs) (listed in Table I). These BMGs cover many typical alloy systems, including Zr-, Cu-, Ca-, Mg-, Ni-, Fe-, and rare earth elements based BMGs, and their thermal, mechanical and physical properties are markedly different (Their values of $T_g$ , $E$ and Poisson's ratio span from 317 K to 930 K, 23 GPa to 195 GPa, 0.276 to 0.41, respectively) [12-13]. The linear fitting results are $E \propto (0.206\pm0.017)T_g$ and $G \propto (0.0759\pm0.0065)T_g$. The relations between $T_g$ and $G$ (and $E$) are key evidence for the elastic model [1], however, it can be seen that the data are rather scattered, and the adjusted R-square (the adjusted coefficients of determination [19]) are 0.755 and 0.741, respectively, which denotes that $G$ or $K$ should not be the only parameter that involved in glass transition.

In fact, according to shoving model, a characteristic volume $V_c$ is involved in homogeneous flow of glass-forming liquids as [1,14]: $\Delta F= GV_c$. In inhomogeneous flow of glasses, the activation energy of a flow event unit (shear transformation zone, STZ) also correlates with volume as [15-16]: $\Delta F = (8/\pi^2)G\gamma_C{}^2\zeta\Omega$, where $\Omega$ is the volume of STZ and $\gamma_C$ is the shear strain limit. Recently, it is found that the elastic moduli scaled with $V_m$ show better correlations with the thermal and mechanical properties for metallic glasses [17-18, 20-21]. Thus, the characteristic volume could be an important parameter involved in the flow event in glass transition and glass. Figure 1(b) shows the plot of $GV_m$ (and $EV_m$) versus $T_g$ for these BMGs. Indeed, the data can be better fitted with $EV_m \propto (1.53\pm0.06)T_g$ and $GV_m \propto (0.560\pm0.029)T_g$, and the adjusted-R increases to 0.923 and 0.905, respectively. The results certify that the combination of $G$ and $V_m$ ($E$ and $V_m$) does show much better correlation with $T_g$ and the $V_m$ is another important parameter in governing the glass transition.

It is found that the atom number $N$ involved in the cooperative flow events of STZ in metallic glasses is similar and around ~100 [15-16], and the volume of STZ can be expressed as $V = \sum_{i=1}^{N\sim100} v_i = N\frac{V_m}{N_A}$ (where $N_A$ is the Avogadro constant). The inhomogeneous flow in glass is a self-organized of a large number of local shearing events (or STZ) [15-17], and the transition from local shearing to macroscopic shear band results from the dramatic increase of the atom mobility and softening along a shear plane motivated by the input mechanical energy [15-17]. Thus, the transition is akin to a process of stress driven glass-to-liquid transition or glass transition [17]. Then, the involved activation energy $GV= NGV_m/N_A$ should have linear correlation with $k_BT_g$. The good linear correlation between $GV_m$ (and $EV_m$) and $T_g$ in Fig. 1(b) confirms that the $N$ involved in the cooperative flow event for various metallic glasses is almost the same.

Based on above scaling laws and elastic model, we propose that it is the flow activation energy density ($\rho_E$), not the flow activation energy itself, correlates with the elastic moduli as:

$$\rho_E = \frac{\Delta F}{V_m} \propto \text{Moduli} \qquad (1)$$

The extended elastic model means that the energy per volume needed in glass transition or in STZ in metallic glass is proportional to the elastic moduli.

Previous elastic models [1] suggest that the atoms or atomic groups go through pure shearing displacement which is independent of $K$, and the $\rho_E$ depends only on $G$. Recent works [18, 20] and the jamming model of granular systems [22] find that both shear and dilatation are involved in the flow during glass transition and deformation. Next, we further justify the flow activation energy density $\rho_E$ relates not to $G$ or $K$ but both $G$ or $K$. That is, the flow event relate to both shearing transformation (corresponding to volume-preserving $G$) and dilatation (corresponding to volume-nonpreserving $K$). At $T_g$, the $\Delta F/T_g$ should be the same for all glasses that is independent of Poisson's ratio or other moduli [3]. However, the statistics analysis of both $GV_m/T_g$ and $KV_m/T_g$ for glasses listed in Table I show linearly depends on $\nu$ as: $KV_m/T_g \propto 8.78\nu$ and $GV_m/T_g \propto -0.86\nu$, respectively [Fig. 2(a)-(b)]. The relationships between $\Delta F/T_g$ and $\nu$ should be neither the dashed black line ($KV_m/T_g$) nor the short-dashed olive line ($GV_m/T_g$) but a constant like the solid magenta line in Fig. 3(a). The slope in the relationship of $KV_m/T_g$ vs $\nu$ is positive, while that of $GV_m/T_g$ vs $\nu$ is negative. The bigger the slope of $KV_m/T_g$ vs $\nu$ (or $GV_m/T_g$ vs $\nu$) is, the less the contribution to $\Delta F$ of the modulus should be. This indicates that if $\Delta F/T_g$ would be a constant it should not only relate to $G$ or $K$ but a combination of both $K$ and $G$. When the ratio of the contribution of $G$ and $K$ is about 8.78: 0.86≈10:1, or alternatively $\rho_E = \Delta F/V_m = (10G+K)/11$, the $\Delta F/T_g$ vs $\nu$ is a constant as shown in Fig. 3(a). Figure 3(b) shows the results of $(0.91G + 0.09K)V_m/T_g$ vs. $\nu$ for various BMGs. The data indeed can be well fitted by a constant of 0.63 and is independent of $\nu$, which is consistent with the glass transition phenomenon that the viscosity of all the liquids gets to the same value at $T_g$. The $\Delta F = (0.91G + 0.09K)V_m$, which is independent of mass or amount, is a kind of elastic energy, and the ratio $\Delta F/RT_g = 0.63/R = 0.076$, which is dimensionless, can be regarded as some kind of elastic strain stored in glassy state [23]. Thus, the glass transition could be regarded as the release or absorb of the elastic strain stored.

The acoustic velocities behaviors during glass transition further verify the extended elastic model. The $T$-dependent transversal ($V_s$) and longitudinal ($V_L$) velocities change differently during the glass transition process [1, 24-25], and the ratio of their relative changes is about: $\frac{\Delta V_S}{V_S} : \frac{\Delta V_L}{V_L} \approx 2:1$. From $\rho V_S^2 = G$ and $\rho V_L^2 = \frac{4}{3}G + K$, we obtain the relative changes of $G$ and $K$ in metallic glasses is: $\frac{\Delta G}{G} : \frac{\Delta K}{K} \approx 5:1$. In 3$D$ space, there are two shear models (corresponding to $G$) and one radial model (dilatation model corresponding to $K$) when atoms move. Thus, the contribution of $G$ should be doubled, and the ratio of the contribution of $G$ and $K$ in $\rho_E$ should be about 10:1, that is: $\rho_E = (10G+K)/11$.

We further discuss why the elastic moduli show better correlation with $\rho_E$ rather than the activation energy itself. The shear elastic energy density $\phi$ of a STZ can be expressed as $\phi(\gamma) = \phi_0 \sin^2(\pi\gamma / 4\gamma_C)$ [15], where $\phi_0$ is the total barrier energy density and $\gamma$ is the shear strain. The $G$ can be reduced from the $\phi$, not the shear elastic energy, in a way of $G = \phi''|_{\gamma=0} = \frac{\phi_0}{8\gamma_C^2/\pi^2}$ [15]. This indicates that $G$ is related to the barrier energy density. The $K$ can be expressed as $K = V_0 \frac{\partial^2 U}{\partial V^2}\Big|_{V=V_0} = \partial^2\left(U/V_0\right)/\partial\left(V/V_0\right)^2\Big|_{V=V_0}$ [21], where $U$ is the atomic potential energy, $V_0$ is the atomic equilibrium volume, $U/V_0$ is the energy density and $V/V_0$ correlates with elastic strain. In harmonic approximation around $V_0$, the $U$ can be expressed in a parabolic form as $U=U_0(1-\alpha V/V_0)^2$ [1], where $\alpha$ is constant depending on the atomic bonding nature. This gives $K= 2\alpha^2 U_0/V_0$, which correlates with the potential energy density at equilibrium state. Thus, both $G$ and $K$ are proportional to their corresponding deformation energy density. Therefore, it is reasonable for $\rho_E$ rather than the activation energy shows correlation with the combination of $K$ and $G$.

Most models for flow in glasses and supercooled liquids consider the case of simple shear, which involves only shear stress and shear modulus. Our model suggests that the both homogeneous and inhomogeneous flows at one hand is shearing process and on the other hand must generate free volume which is a form of dilatation (In fact, the shear induced dilatation has been widely observed [22]), and demonstrates that both shear and free volume are important for flow in glass transition and deformation, and provides an intuitional picture of the flow of the atoms or atomic groups in glass or liquid. Furthermore, the formation of shear bands when the BMGs deform plastically is thought to be akin to the process of glass transition [17, 23]. Thus, this means that both the shear [26] and the dilatation [22] could be involved in the formation of shear bands. However, due to the critical difference between the two phenomena: the glass transition is constraint-free, while the formation of shear bands is stress-constraint, and the formation of shear bands then may involve less dilatation.

**Acknowledgements:** The experimental assistance of DQ Zhao and RJ Wang is appreciated. Financial support is from the NSFC (Nrs. 50731008 and 50921091) and MOST 973 (No. 2007CB613904 and 2010CB731603).

Table I. The compositions, $T_g$, average molar volume $V_m$, $K$, $E$, $G$, $\nu$, and the combined parameters $Moduli{\cdot}V_m/T_g$ of 46 different kinds of BMGs [11-21].

| Compositions | $T_g$ K | $V_m$ cm$^3$/mol | $K$ GPa | $E$ GPa | $G$ GPa | $\nu$ | $GV_m/T_g$ | $KV_m/T_g$ | $(0.91G + 0.09K)V_m/T_g$ |
|---|---|---|---|---|---|---|---|---|---|
| $Ca_{65}Mg_{8.54}Li_{9.96}Zn_{16.5}$ | 317 | 20.25 | 20.1 | 23.4 | 8.9 | 0.307 | 0.572 | 1.287 | 0.636 |
| $Ca_{65}Mg_{8.31}Li_{9.69}Zn_{17}$ | 320 | 20.10 | 18.4 | 23.2 | 9.0 | 0.291 | 0.564 | 1.159 | 0.618 |
| $Yb_{62.5}Zn_{15}Mg_{17.5}Cu_5$ | 385 | 19.24 | 19.8 | 26.5 | 10.4 | 0.276 | 0.520 | 0.989 | 0.562 |
| $Ce_{70}Al_{10}Ni_{10}Cu_{10}$ | 359 | 16.94 | 27.0 | 30.3 | 11.5 | 0.314 | 0.543 | 1.274 | 0.609 |
| $(Ce_{20}La_{80})_{68}Al_{10}Cu_{20}Co_2$ | 366 | 16.78 | 32.6 | 31.8 | 11.9 | 0.338 | 0.544 | 1.496 | 0.629 |
| $Ce_{68}Al_{10}Cu_{20}Nb_2$ | 345 | 16.70 | 30.1 | 31.0 | 11.7 | 0.328 | 0.564 | 1.455 | 0.644 |
| $(Ce_{80}La_{20})_{68}Al_{10}Cu_{20}Co_2$ | 355 | 16.69 | 31.8 | 31.1 | 11.6 | 0.337 | 0.547 | 1.494 | 0.632 |
| $Ce_{68}Al_{10}Cu_{20}Co_2$ | 352 | 16.57 | 30.3 | 31.3 | 11.8 | 0.328 | 0.555 | 1.428 | 0.634 |
| $Ce_{68}Al_{10}Cu_{20}Ni_2$ | 352 | 16.57 | 31.8 | 31.9 | 12.0 | 0.333 | 0.564 | 1.495 | 0.648 |
| $Ce_{68}Al_{10}Cu_{20}Co_2$ | 351 | 16.44 | 30.1 | 30.3 | 11.5 | 0.333 | 0.532 | 1.411 | 0.611 |
| $La_{60}Al_{20}Co_{20}$ | 477 | 15.96 | 39.2 | 38.7 | 14.5 | 0.335 | 0.486 | 1.311 | 0.560 |
| $Pr_{55}Al_{25}Co_{20}$ | 509 | 15.07 | 43.5 | 45.9 | 15.4 | 0.341 | 0.456 | 1.287 | 0.531 |
| $Dy_{55}Al_{25}Co_{20}$ | 635 | 14.27 | 52.2 | 61.4 | 23.5 | 0.304 | 0.529 | 1.174 | 0.587 |
| $Tb_{55}Al_{25}Co_{20}$ | 612 | 14.15 | 50.2 | 59.5 | 22. 9 | 0.302 | 0.528 | 1.160 | 0.585 |
| $Ho_{55}Al_{25}Co_{20}$ | 649 | 13.85 | 58.8 | 66.6 | 25.4 | 0.311 | 0.542 | 1.255 | 0.607 |
| $Er_{55}Al_{25}Co_{20}$ | 663 | 13.55 | 60.7 | 70.7 | 27.1 | 0.306 | 0.553 | 1.241 | 0.615 |
| $Tm_{39}Y_{16}Al_{25}Co_{20}$ | 664 | 13.51 | 66.1 | 77.5 | 29.7 | 0.305 | 0.604 | 1.345 | 0.671 |
| $Tm_{55}Al_{25}Co_{20}$ | 678 | 13.47 | 62.0 | 72.2 | 25.6 | 0.319 | 0.509 | 1.232 | 0.574 |
| $Lu_{39}Y_{16}Al_{25}Co_{20}$ | 687 | 13.30 | 71.3 | 78.9 | 30.0 | 0.316 | 0.581 | 1.380 | 0.653 |
| $Lu_{45}Y_{10}Al_{25}Co_{20}$ | 698 | 13.25 | 70.2 | 79.1 | 31.1 | 0.307 | 0.590 | 1.332 | 0.657 |
| $Lu_{55}Al_{25}Co_{20}$ | 701 | 13.20 | 69.2 | 80.0 | 30.6 | 0.307 | 0.576 | 1.303 | 0.642 |
| $Mg_{65}Cu_{25}Gd_{10}$ | 421 | 12.51 | 45.1 | 50.6 | 19.3 | 0.313 | 0.573 | 1.340 | 0.642 |
| $Mg_{65}Cu_{25}Y_9Gd_1$ | 423 | 12.37 | 39.0 | 52.2 | 20.4 | 0.277 | 0.597 | 1.142 | 0.646 |
| $Mg_{65}Cu_{25}Y_{10}$ | 419 | 12.36 | 41.4 | 49.1 | 18.9 | 0.302 | 0.556 | 1.220 | 0.616 |
| $Mg_{65}Cu_{25}Y_8Gd_2$ | 420 | 12.23 | 39.9 | 51.7 | 20.1 | 0.284 | 0.586 | 1.161 | 0.638 |
| $Mg_{65}Cu_{25}Y_5Gd_5$ | 422 | 12.05 | 39.1 | 50.6 | 19.7 | 0.284 | 0.563 | 1.117 | 0.613 |
| $Mg_{65}Cu_{25}Tb_{10}$ | 415 | 11.95 | 44.7 | 51.3 | 19.6 | 0.309 | 0.565 | 1.288 | 0.630 |
| $Zr_{64.13}Cu_{15.75}Ni_{10.12}Al_{10}$ | 640 | 11.68 | 106.6 | 78.4 | 28.5 | 0.377 | 0.519 | 1.946 | 0.648 |
| $Zr_{65}Cu_{15}Ni_{10}Al_{10}$ | 652 | 11.65 | 106.7 | 83.0 | 30.3 | 0.37 | 0.541 | 1.906 | 0.664 |
| $Zr_{61.88}Cu_{18}Ni_{10.12}Al_{10}$ | 651 | 11.51 | 108.3 | 80.1 | 29.1 | 0.377 | 0.514 | 1.915 | 0.640 |
| $Zr_{55}Al_{19}Co_{19}Cu_7$ | 733 | 11.44 | 114.9 | 101.7 | 30.8 | 0.377 | 0.481 | 1.794 | 0.599 |
| $Zr_{57}Nb_5Cu_{15.4}Ni_{12.6}Al_{10}$ | 687 | 11.44 | 107.7 | 87.3 | 32.0 | 0.365 | 0.533 | 1.793 | 0.646 |
| $Zr_{57}Ti_5Cu_{20}Ni_8Al_{10}$ | 657 | 11.43 | 99.2 | 82.0 | 30.1 | 0.362 | 0.523 | 1.725 | 0.632 |
| $(Zr_{59}Ti_6Cu_{22}Ni_{13})_{85.7}Al_{14.3}$ | 689 | 10.74 | 112.6 | 92.7 | 34.0 | 0.363 | 0.530 | 1.755 | 0.640 |
| $Cu_{45}Zr_{45}Al_7Gd_3$ | 670 | 10.71 | 105.9 | 90.1 | 33.2 | 0.358 | 0.530 | 1.692 | 0.635 |
| $Zr_{46.75}Ti_{8.25}Cu_{10.15}Ni_{10}Be_{27.25}$ | 622 | 10.21 | 111.9 | 100 | 37.2 | 0.35 | 0.610 | 1.836 | 0.721 |
| $Zr_{48}Nb_8Cu_{12}Fe_8Be_{24}$ | 658 | 10.17 | 113.6 | 95.7 | 35.2 | 0.36 | 0.544 | 1.756 | 0.653 |
| $Zr_{41}Ti_{14}Cu_{12.5}Ni_{10}Be_{22.5}$ | 625 | 9.79 | 114.1 | 101 | 37.4 | 0.352 | 0.586 | 1.787 | 0.694 |
| $Ni_{45}Ti_{20}Zr_{25}Al_{10}$ | 733 | 9.61 | 129.6 | 109 | 40.2 | 0.359 | 0.527 | 1.699 | 0.632 |
| $Cu_{60}Zr_{20}Hf_{10}Ti_{10}$ | 754 | 9.50 | 128.2 | 101 | 36.9 | 0.369 | 0.465 | 1.616 | 0.569 |
| $Pd_{77.5}Cu_6Si_{16.5}$ | 633 | 8.74 | 166.0 | 89.7 | 31.8 | 0.41 | 0.439 | 2.293 | 0.606 |
| $Pd_{64}Ni_{16}P_{20}$ | 630 | 8.29 | 166.0 | 91.9 | 32.7 | 0.408 | 0.430 | 2.183 | 0.588 |
| $Pd_{40}Cu_{40}P_{20}$ | 590 | 7.98 | 158.0 | 93.0 | 33.2 | 0.402 | 0.449 | 2.136 | 0.601 |
| $Pd_{39}Ni_{10}Cu_{30}P_{21}$ | 560 | 7.97 | 159.1 | 98.2 | 35.1 | 0.397 | 0.500 | 2.264 | 0.658 |
| $Fe_{53}Cr_{15}Mo_{14}Er_1C_{15}B_6$ | 900 | 7.94 | 180.0 | 195 | 75.0 | 0.317 | 0.610 | 1.588 | 0.698 |
| $Fe_{61}Mn_{10}Cr_4Mo_6Er_1C_{15}B_6$ | 930 | 7.48 | 146.0 | 193 | 75.0 | 0.281 | 0.603 | 1.174 | 0.654 |

## Figure captions

Figure 1. (Color online) (a) The Young's modulus $E$ and shear modulus $G$ of 46 kinds of metallic glasses versus $T_g$. (b) The combination of moduli and molar volume $V_m$ versus $T_g$ follow a better relationship. The solid lines are the linear fitting of the experimental data.

Figure 2. (Color online) (a) The $KV_m/T_g$ versus Poisson's ratio $\nu$ and (b) $GV_m/T_g$ versus $\nu$ for 46 kinds of metallic glasses. The lines are the linear fit of the experiment data.

Figure 3. (Color online) (a) The $Moduli \cdot V_m/T_g$ versus $\nu$. The black dashed line represents for $KV_m/T_g$ vs. $\nu$, the dark cyan short dashed line for $GV_m/T_g$ vs. $\nu$, and the red solid line for $\rho_E$ =*Moduli* which is Poisson's ratio independent. (b) The experiment data of $(0.91G+0.09K)V_m/T_g$ versus $\nu$ are well linearly fitted, denoting that the $(0.91G+0.09K)V_m/T_g$ for various metallic glasses is independent of $\nu$.

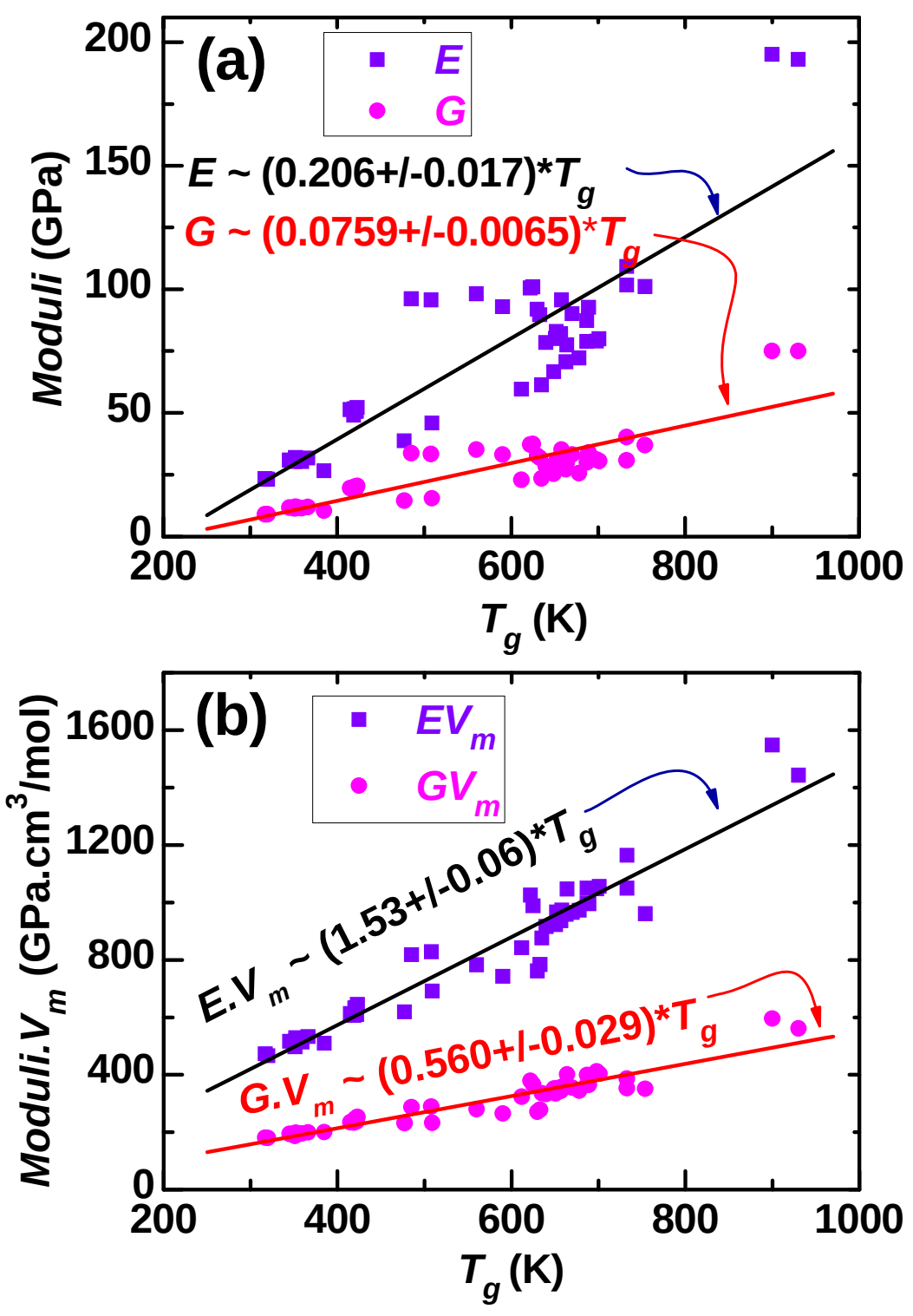

Figure 1. J. Q. Wang, et al.

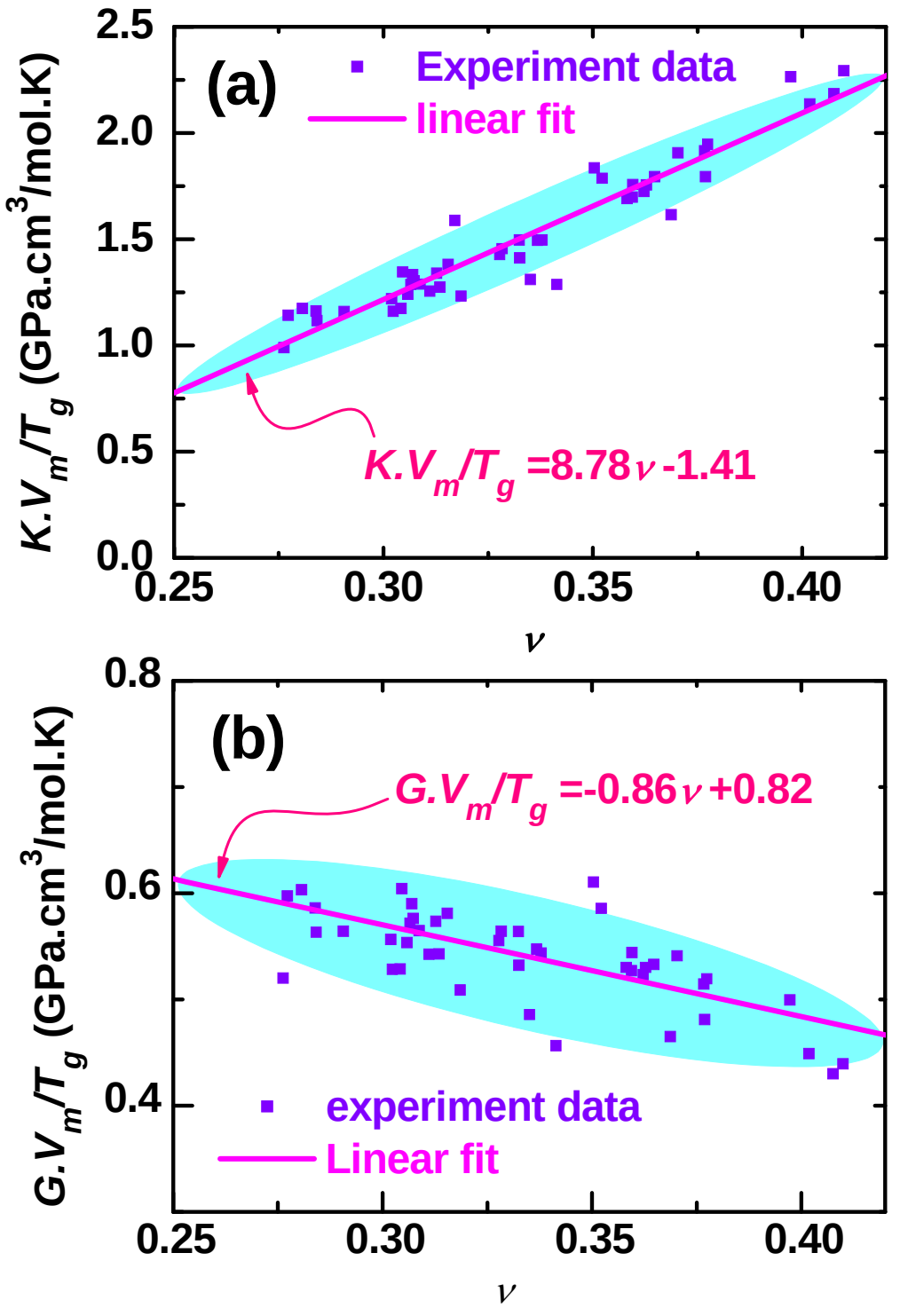

Figure 2. J. Q. Wang, et al.

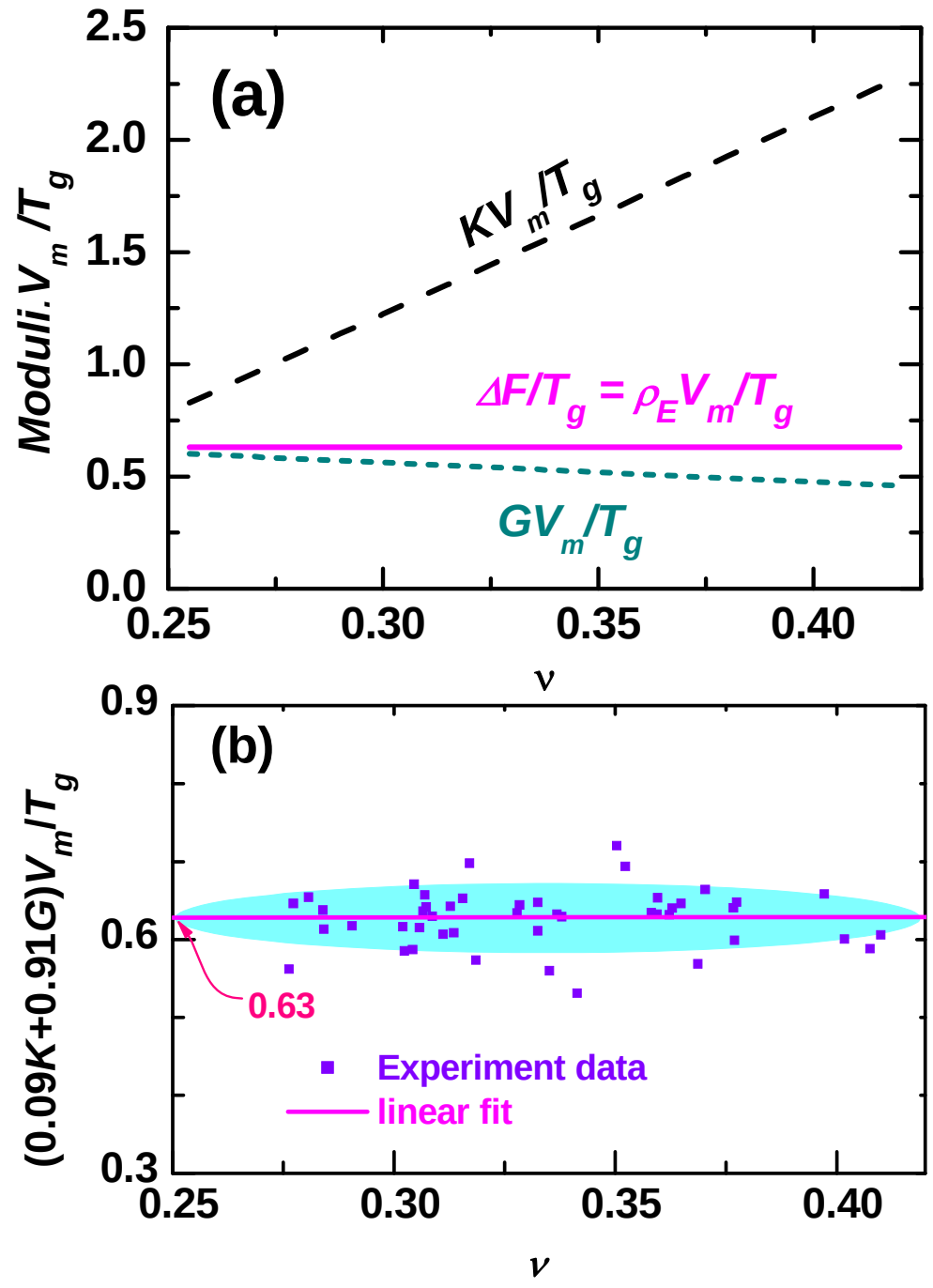

Figure 3. J. Q. Wang, et al.